\documentclass[12pt]{article}
\usepackage{hyperref}
\usepackage{amsmath,amssymb}
\usepackage{url,color}
\usepackage{graphicx}
\usepackage{slashed}
\usepackage{multirow} % more control on table columns/rows
\usepackage{booktabs} % for better looking tables
\usepackage{pifont} % checkmarks and crosses
\makeatletter
\newif\if@preliminary
\@preliminaryfalse
\def\preliminary{\@preliminarytrue}

\def\preprintno#1{\def\@preprintno{#1}}
\def\address#1{\def\@address{#1}}
\def\email#1#2{\thanks{\tt #1@{}#2}}
\def\abstract#1{\def\@abstract{#1}}
\renewcommand\abstractname{ABSTRACT}
\newlength\preprintnoskip
\newlength\abstractwidth
\@titlepagetrue
\renewcommand\maketitle{\begin{titlepage}%
  \let\footnotesize\small
  \hfill\parbox{\preprintnoskip}{%
  \begin{flushright}\@preprintno\end{flushright}}\hspace*{1cm}
  \vskip 60\p@
  \begin{center}%
    {\Large\bf\boldmath \@title \par}\vskip 1cm%
    {\sc\@author \par}\vskip 3mm%
    {\@address \par}%
    \if@preliminary
      \vskip 2cm {\large\sf PRELIMINARY DRAFT \par \@date}%
    \fi
  \end{center}\par
  \@thanks
  \vfill
  \begin{center}%
    \parbox{\abstractwidth}{\centerline{\abstractname}%
    \vskip 3mm%
    \@abstract}
  \end{center}
  \end{titlepage}%
  \setcounter{footnote}{0}%
  \let\thanks\relax\let\maketitle\relax
  \gdef\@thanks{}\gdef\@author{}\gdef\@address{}%
  \gdef\@title{}\gdef\@abstract{}\gdef\@preprintno{}
}%
\def\@citex[#1]#2{\if@filesw\immediate\write\@auxout{\string\citation{#2}}\fi
  \def\@citea{}\@cite{\@for\@citeb:=#2\do
    {\@citea\def\@citea{,\penalty\@m}\@ifundefined
       {b@\@citeb}{{\bf ?}\@warning
       {Citation `\@citeb' on page \thepage \space undefined}}%
\hbox{\csname b@\@citeb\endcsname}}}{#1}}
\def\citerange{\@ifnextchar [{\@tempswatrue\@citexr}{\@tempswafalse\@citexr[]}}
\def\@citexr[#1]#2{\if@filesw\immediate\write\@auxout{\string\citation{#2}}\fi
  \def\@citea{}\@cite{\@for\@citeb:=#2\do
    {\@citea\def\@citea{--\penalty\@m}\@ifundefined
       {b@\@citeb}{{\bf ?}\@warning
       {Citation `\@citeb' on page \thepage \space undefined}}%
\hbox{\csname b@\@citeb\endcsname}}}{#1}}
\long\def\@makecaption#1#2{%
  \vskip\abovecaptionskip
  \sbox\@tempboxa{#1: \emph{#2}}%
  \ifdim \wd\@tempboxa >\hsize
    #1: \emph{#2}\par
  \else
    \hbox to\hsize{\hfil\box\@tempboxa\hfil}%
  \fi
  \vskip\belowcaptionskip}
\makeatother
\newcommand{\ii}{\mathrm{i}}

\newcommand{\pd}{\partial}
\newcommand{\LL}{\mathcal{L}}
\newcommand{\eff}{{\rm eff}}
\newcommand{\vV}{\mathbf{V}}
\newcommand{\vT}{\mathbf{T}}
\newcommand{\vW}{\mathbf{W}}
\newcommand{\vw}{\mathbf{w}}

\newcommand{\tr}[1]{\operatorname{tr}\left[#1\right]}
\newcommand{\pp}{{\prime\,2}}
\newcommand{\vB}{\mathbf{B}}
\newcommand{\vD}{\mathbf{D}}
\newcommand{\sw}{s_{\mathrm{w}}}
\newcommand{\cw}{c_{\mathrm{w}}}

\newcommand{\GeV}{\text{GeV}}
\newcommand{\TeV}{\text{TeV}}

\newcommand{\pb}{\text{pb}}
\newcommand{\amp}{\mathcal{A}}
\newcommand{\bss}{\begin{tiny}}
\newcommand{\ess}{\end{tiny}}
\begin{document}

%begin{fmffile}{wopics}
%begin{empfile}

%%%%%%%%%%%%%%%%%%%%%%%%%%%%%%%%%%%%%%%%%%%%%%%%%%%%%%%%%%%%%%%%%%%%%%%%
\preprintno{%
DESY 13-132 \\ 
SI-HEP-2013-06}

\title{%
  Simplified Models for New Physics in Vector Boson Scattering --
  Input for Snowmass 2013}

\author{
  J\"urgen Reuter\email{juergen.reuter}{desy.de}$^a$, 
  Wolfgang Kilian\email{kilian}{physik.uni-siegen.de}$^b$, 
  Marco Sekulla\email{sekulla}{physik.uni-siegen.de}$^b$}

\address{\it%
    $^a$DESY Theory Group,
    D--22603 Hamburg, Germany\\
    $^b$University of Siegen, Department of Physics, 
    D--57068 Siegen, Germany
}

\date{%
  \today\\
  \hfil\\
  \texttt{$ $Id: 2013_Snowmass_Resonance.tex 1 2013-07-12 19:35:11Z
    jr_reuter $ $}} 

\abstract{%
In this contribution to the Snowmass process 2013 (which is a
preliminary version of~\cite{KRS}) we give a brief
review of how new physics could enter in the electroweak (EW) sector 
of the Standard Model (SM). This new physics, if it is directly
accessible at low energies, can be parameterized by explicit
resonances having certain quantum numbers. The extreme case is the
decoupling limit where those resonances are very heavy and leave
only traces in form of deviations in the SM
couplings. Translations are given into higher-dimensional operators
leading to such deviations. As long as such resonances are
introduced without a UV-complete theory behind it, these models
suffer from unitarity violation of perturbative scattering
amplitudes. We show explicitly how theoretically sane descriptions
could be achieved by using a unitarization prescription that allows
a correct description of such a resonance without specifying a
UV-complete model. 
}
\maketitle

\section{Introduction: extended electroweak symmetry breaking}
\label{sec:intro}

After the first run of LHC, a light Higgs particle compatible with
the predictions from the SM and electroweak precision tests
(EWPT) with a mass of $m_H = 125$ GeV has been found. At the moment,
all production cross sections and decay rates are compatible with the
SM. This, however, is not so astonishing as it is rather difficult to
cook up a BSM model that is compatible with EWPT and has deviations in
the Higgs properties larger than the experimental uncertainties of the
2011/12 data. So, we got a glimpse on the electroweak symmetry
breaking mechanism, but not yet the full answer. We have to check
whether longitudinal electroweak vector boson scattering,
i.e. scattering of Higgs field components really behaves in the way as
expected from the SM. Furthermore we have to look at possible
additional states that couple to the electroweak system of the $W$,
$Z$ and the 125 GeV state. Almost all BSM models predict modifications
of the EW sector as part of the explanations for the hierarchy problem,
namely the stability of a fundamental scalar state under radiative
corrections. Examples are extra-dimensional models that comprise
Kaluza-Klein recurrences of the EW gauge bosons and possibly
also the Higgs, SUSY models as more generally two- or multi-doublet
Higgs models, Little Higgs models, Twin Higgs models, models of
complete or partial compositeness, Technicolor- or Topcolor-like
models etc.

Usually, a fine-tuning measure is used for the definition of the
little hierarchy problem: if the parameters of a model have to be
tuned to a higher precision in order to achieve the correct Higgs mass
parameter, the higher the fine tuning is. Though this is not a
physical argument per se, it might give a guideline how contrived a
model is. Most models seem the most natural for extensions of the
electroweak sector ``just around the corner'', i.e. very close to the
EW scale itself (as a prime example, this has been analyzed for Little
Higgs models quite recently~\cite{LHM}). 

Any model that is believed to solve the hierarchy problem is endowed
with some sort of sector of new physics that couples to the EW
bosons. The goal of this contribution to the Snowmass white paper is
to define a Simplified Model that is able to describe the essence of
this new physics sector in an approach as model-independent way as
possible. We refrain from discussing fermionic resonances here, as
those would contribute only at the 1-loop order to vector boson
scattering, and concentrate on new bosonic resonances. To do so, one
needs to supplement the Lagrangian of the EW SM (accounting for the
discovery of the 125 GeV state as the SM Higgs boson but maybe
allowing its couplings to deviate within the limits of the EWPT from
their SM values). As the main signatures to study the EW sector of the
SM are diboson, triboson and generically multi-boson production as
well as vector-boson scattering (VBS), and here particularly
scattering of longitudinal gauge bosons, it is convenient to use an
operator basis containing explicitly the longitudinal degrees of
freedom (DOFs) of the EW gauge bosons. This effective Lagrangian is
basically identical to the chiral EW Lagrangian~\cite{EWLag}, except
that we linearize the Lagrangian by adding the Higgs particle, and all
higher-dimensional operators stem from BSM contributions. So we
implement $SU(2)_L \times U(1)_Y$ gauge invariance, where the building
blocks are the SM fermions, $\psi$, the EW (transversal) gauge boson
fields $W_\mu^a \; (a=1,2,3)$ and $B_\mu$ as well as the longitudinal
DOFs, $\Sigma = \exp \left[ \frac{- i}{v} w^a \tau^a \right]$. Our
first goal is to write down the minimal EW Lagrangian, and add then
deviations from that Lagrangian in the form of higher-dimensional
operators allowed by gauge symmetry as well as CP. Later, we show how
precisely such couplings can arise when heavy BSM resonances are in
the game. The minimal Lagrangian including gauge interactions is then:
\begin{multline}
  \LL_{\text{min}} = \sum_\psi \overline{\psi} (i \slashed{D})
  \psi - \sum_\psi \overline{\psi}_L
        \Sigma M \psi_R - \frac{1}{2 g^2} \tr{\vW_{\mu\nu} \vW^{\mu\nu}} - 
  \frac{1}{2 g^\pp} \tr{\vB_{\mu\nu} \vB^{\mu\nu}} \\ +
   \left (\partial_\mu \phi \right)^\dagger \partial^\mu \phi + V ( \phi ) +
   \frac{g_h v}{2} \tr{\mathbf{V}^\mu \mathbf{V}_\mu} h 
   - \sum_\psi \overline{\psi}_L M \psi_R \phi  
   + \frac{v^2}{4} \tr{(\vD_\mu \Sigma) (\vD^\mu \Sigma)} 
\end{multline}
Here, bold-faced quantities are always in the vector representation of
$SU(2)_L$, $D$ is the corresponding gauge-covariant derivative. $ \phi
= \frac{1}{\sqrt{2}} ( 0, v + h)^T$ is the field of the Higgs
particle
% , and $D_\mu \phi = (\partial_\mu+\mathbf{V}_\mu) \phi$ is the
% gauge-covariant derivative of the Higgs field.
$\vV$ is the field of
longitudinal vectors, $\vV = \Sigma (\vD \Sigma)^\dagger$ that will be
used shortly to write down operators giving rise to modified
couplings. $V(\phi)$ contains the trilinear and quadrilinear Higgs
self-couplings as well as the Higgs mass term. In order to write down
operators projecting out the neutral component, one uses $\vT = \Sigma
\tau^3 \Sigma^\dagger$.  

There are two extreme limits. One is the unitary gauge where 
the Goldstone fields are rotated away: $\vw\equiv 0$,
i.e., $\Sigma\equiv 1$. Here, $\vV \longrightarrow 
-\frac{\ii g}{2}\left[\sqrt2(W^+\tau^+ + W^-\tau^-) + \frac1\cw Z
\tau^3\right]$ and $\vT \longrightarrow \tau^3$. The other
is the gaugeless limit, which removes the transversal DOFs by $g,g'
\to 0$. This limit makes calculations for scattering processes of
longitudinal gauge bosons particularly simple, and was the choice
within approaches for Higgsless models or models with strongly
interacting $Ws$ and/or (very) heavy Higgs bosons. Here, one has $\vV
\longrightarrow \frac{\ii}{v} \bigg\{\: \sqrt2\pd w^+ \tau^+
+\sqrt2\pd w^- \tau^- + \pd z \tau^3 \bigg\} + O(v^{-2})$ and
$\vT \longrightarrow \tau^3 + 2\sqrt2\frac{\ii}{v}\left(w^+\tau^+ -
w^-\tau^-\right) + O(v^{-2})$. 

This minimal (SM) Lagranigan can now be supplemented by additional
operators,  
\begin{equation}
  \LL_\eff = \LL_{\text{min}}  + \beta_1 \LL'_0 + \sum_i \alpha_i \LL_i + 
  \frac{1}{\Lambda} \sum_i \alpha_i^{(5)} \LL^{(5)} +  
  \frac{1}{\Lambda^2} \sum_i \alpha_i^{(6)} \LL^{(6)} + \ldots 
\end{equation}
where $\Lambda$ is (up to $\mathcal{O}(1)$ constants) the scale where
BSM physics potentially enters.
%%%\begin{equation}
\begin{xalignat*}{2}
  \label{operators}
  \LL'_0 &=\; \frac{v^2}{4} \tr{\vT \vV_\mu} \tr{\vT \vV^\mu} & &
  \\
  \LL_1 &=\; \tr{\vB_{\mu\nu} \vW^{\mu\nu}} 
  &
  \LL_6 &=\; \tr{\vV_\mu \vV_\nu} \tr{\vT \vV^\mu} \tr{\vT
    \vV^\nu}
  \\
  \LL_2 &=\; \ii \tr{\vB_{\mu\nu} \lbrack \vV^\mu , \vV^\nu
    \rbrack}
  & 
  \LL_7 &=\; \tr{\vV_\mu \vV^\mu} \tr{\vT \vV_\nu} \tr{\vT
    \vV^\nu}
  \\
  \LL_3 &=\; \ii \tr{\vW_{\mu\nu} \lbrack \vV^\mu , \vV^\nu
    \rbrack} 
  &
  \LL_8 &=\; \tfrac14 \tr{\vT \vW_{\mu\nu}} \tr{\vT \vW^{\mu\nu}} 
  \\
  \LL_4 &=\; \tr{\vV_\mu \vV_\nu} \tr{\vV^\mu \vV^\nu}
  &
  \LL_9 &=\; \tfrac{\ii}{2} \tr{\vT \vW_{\mu\nu}} \tr{\vT \lbrack
    \vV^\mu , \vV^\nu \rbrack}
  \\
  \LL_5 &=\; \tr{\vV_\mu \vV^\mu} \tr{\vV_\nu \vV^\nu}
  &
  \LL_{10} &=\; \tfrac12 \left( \tr{\vT \vV_\mu} \tr{\vT \vV^\mu}
  \right)^2 
\end{xalignat*}
%%%\end{equation}
For more technical details about this formalism interpreted in that
context of simplified models for extended EW symmetry breaking,
cf.~\cite{Alboteanu:2008my,KRS}. Indirect information on new physics
is encoded in the operator coefficients $\beta_1$, $\alpha_i$. 

From EWPT (SLC/LEP/Tevatron measurements) one knows that $\alpha_i
\ll 1$, while on the other hand from naive dimensional analysis one
would assume $\alpha_i \sim 1/16\pi^2 \approx 0.006$ as they have to
renormalize divergencies in an effective field theoretic simplified
model of a UV-complete BSM model. Using such a bottom-up approach
is notoriously difficult. The usual setup has a ratio of the EW and
the BSM scale $\alpha_i = v^2/\Lambda^2$, which is only valid upto unknown
operator normalization coefficients (that are in general coupling
constants of the UV-complete model). Furthermore, the power counting
can be highly nontrivial, producing unexpected scaling behavior of
operators. 

One way to deal with this in a model-independent is to consider
resonances that couple to EWSB sector, what we will do in the next
section. For completeness, we repeat the formulae for triple and
quartic gauge couplings, and how they depend on the SM parameters as 
well as on the operator coefficients of the effective Lagrangian
above:
  
  \begin{align}
    \LL_{TGC} &=  \ii e\left[
      g_1^\gamma A_\mu \left(W^-_\nu W^{+\mu\nu} - W^+_\nu
        W^{-\mu\nu}\right) 
      + \kappa^\gamma W^-_\mu W^+_\nu A^{\mu\nu}
      + \frac{\lambda^\gamma}{M_W^2}W^-_\mu{}^\nu
      W^+_{\nu\rho} A^{\rho\mu} 
    \right]
    \nonumber\\ &\quad + \ii e\frac{\cw}{\sw}\left[
      g_1^Z Z_\mu \left(W^-_\nu W^{+\mu\nu} - W^+_\nu
        W^{-\mu\nu}\right) 
      + \kappa^Z W^-_\mu W^+_\nu Z^{\mu\nu}
      + \frac{\lambda^Z}{M_W^2}W^-_\mu{}^\nu W^+_{\nu\rho}
      Z^{\rho\mu} 
    \right] \, , \\ 
  \LL_{QGC} &=
  e^2\left[ g_1^{\gamma\gamma} A^\mu A^\nu W^-_\mu W^+_\nu
    -g_2^{\gamma\gamma} A^\mu A_\mu W^{-\nu} W^+_\nu\right]
  \nonumber\\ &\quad
  + e^2\frac{\cw}{\sw}\left[ g_1^{\gamma Z} A^\mu Z^\nu
    \left(W^-_\mu W^+_\nu + W^+_\mu W^-_\nu\right)
    -2g_2^{\gamma Z} A^\mu Z_\mu W^{-\nu} W^+_\nu \right]
  \nonumber\\ &\quad
  + e^2\frac{\cw^2}{\sw^2}\left[ g_1^{ZZ} Z^\mu Z^\nu W^-_\mu W^+_\nu
    -g_2^{ZZ} Z^\mu Z_\mu W^{-\nu} W^+_\nu\right]
  \nonumber\\ &\quad
  + \frac{e^2}{2\sw^2}\left[ g_1^{WW} W^{-\mu} W^{+\nu} W^-_\mu W^+_\nu
    -g_2^{WW}\left(W^{-\mu} W^+_\mu\right)^2\right]
  % \nonumber\\ &\quad
  + \frac{e^2}{4\sw^2\cw^4} h^{ZZ} (Z^\mu Z_\mu)^2
\end{align}

In these equations, the SM values are
$g_1^{\gamma,Z}=\kappa^{\gamma,Z}=1$, $\lambda^{\gamma,Z}=0$, and
$g_{1/2}^{VV'}=1$, $h^{ZZ} = 0$. The quantity $\delta_Z =
\frac{\beta_1+g^\pp\alpha_1}{\cw^2-\sw^2}$ takes into account the
definition of the EW scheme as well as the oblique corrections through
the $\rho$ parameter. In the presence of the operators
Eq.~\ref{operators}, one gets the following shifts: 
\begin{align}
  \Delta g_1^\gamma &= 0
  &
  \Delta\kappa^\gamma &= g^2(\alpha_2-\alpha_1) + g^2\alpha_3 
  + g^2(\alpha_9-\alpha_8)
  \\
  \Delta g_1^Z &= \delta_Z + \tfrac{g^2}{\cw^2}\alpha_3
  &
  \Delta\kappa^Z &= \delta_Z - g^\pp(\alpha_2-\alpha_1) + g^2\alpha_3 
  + g^2(\alpha_9-\alpha_8)
\end{align}
as well as 
\begin{align}
  \Delta g_1^{\gamma\gamma} &= \Delta g_2^{\gamma\gamma} = 0
  &
  \Delta g_2^{ZZ} &= 2\Delta g_1^{\gamma Z}
  - \tfrac{g^2}{\cw^4}(\alpha_5 + \alpha_7)
  \\
  \Delta g_1^{\gamma Z} &= \Delta g_2^{\gamma Z}
  = \delta_Z + \tfrac{g^2}{\cw^2}\alpha_3
  &
  \Delta g_1^{WW} &= 2\cw^2\Delta g_1^{\gamma Z} +
  2g^2(\alpha_9-\alpha_8) 
  + g^2\alpha_4
  \\
  \Delta g_1^{ZZ} &= 2\Delta g_1^{\gamma Z}
  + \tfrac{g^2}{\cw^4}(\alpha_4 + \alpha_6)
  &
  \Delta g_2^{WW} &= 2\cw^2\Delta g_1^{\gamma Z} +
  2g^2(\alpha_9-\alpha_8) 
  - g^2\left(\alpha_4 + 2\alpha_5\right)
  \vphantom{\tfrac{g^2}{\cw^2}}
  \\
  h^{ZZ} &= g^2\left[\alpha_4 + \alpha_5 
    + 2\left(\alpha_6+\alpha_7 + \alpha_{10}\right)\right] 
\end{align}

%%%%% %%%%% %%%%%

\section{Electroweak Resonances and Translation into Anomalous
  Couplings}

It has been discussed how higher-dimensional
operators lead to deviations of triple and quartic gauge
couplings from their SM values in the previous section. Here, we want to describe
how BSM models in their incarnation as EW resonances coupling to the
SM EW gauge boson sector (particularly the longitudinal DOFs) generate
such anomalous couplings. To be as general as possible, we include
weakly interacting cases (e.g. Little Higgs models) where the new
resonances are narrow (proper particles), as well as strongly
interacting cases (e.g. compositeness or Technicolor models) where the
new resonances are rather wide and could even approach the case of a
continuum (e.g. unparticles or conformal sectors). As we know from
EWPT, $\beta_1 \ll 1$, so the $SU(2)_c$ custodial symmetry of weak
isospin (that in the SM is only broken by hypercharge $g' \neq 0$ 
and the fermion masses) is valid to a very good approximation.
Only resonances listed in the following table with spin $J$ and weak
isospin $I$ can couple to a system of two EW vector bosons. 
\vspace{1mm}

\begin{center}
\begin{tabular}{|r|ccc|}\hline
  & 
  $J = 0$ & 
  $J = 1$ & 
  $J = 2$ 
  \\\hline
  $I = 0$ & 
  $\sigma^0 \;\text{(``Higgs'')}$ &
  $[\omega^0] \; (\gamma'/Z')$ &
  $f^0 \; \text{(KK graviton)}$ 
  \\          
  $I = 1$ & 
  $[\pi^\pm, \pi^0] \; \text{(2HDM)}$ & 
  $\rho^\pm, \rho^0 \; (W'/Z')$ & 
  $[a^\pm, a^0]$ 
  \\ 
  $I = 2$ & $\phi^{\pm\pm}, \phi^\pm, \phi^0 \;
  \text{(Higgs triplet)}$ & 
  $\text{---}$ &
  $t^{\pm\pm}, t^\pm, t^0$ 
  \\\hline
\end{tabular}
\end{center}

\vspace{1mm}

The table shows prime examples
for the corresponding combinations where a specific choice for the
hypercharge has been made. The entries in brackets are combinations
that are only possible with $SU(2)_c$-violating couplings, and are not
further discussed here. The scalar isoscalar has the same quantum
numbers as the SM Higgs boson. The scalar isovector appears in
Technicolor models, while the isotensor can be found in the Littlest
Higgs model, e.g. Vector resonances appear in extra-dimensional
models, Technicolor, Little Higgs models and many more. Tensor
resonances without EW quantum numbers can be thought of as a
recurrence of the graviton, while the isovector and -tensor are quite
exotic and appear only e.g. in extended compositeness models. 

As a next step, we relate these resonances from our simplified models
to anomalous couplings. Consider any kind of heavy resonance with
generic Lagrangian $ \LL_\Phi = z \Bigl[ \Phi
\left( M_\Phi^2 + D D \right)\Phi + 2 \Phi J \Bigr]$. Here, $z$ is a
(wavefunction re)normalization constant of the Lagrangian, and $D$ is
the gauge-covariant derivative. $J$ is the EW current to which that
particular resonance couples. Integrating out the resonance leeds to 
$\LL_\Phi^\eff = - \frac{z}{M^2} JJ + \frac{z}{M^4} J (D D) J +
\mathcal{O} (M^{-6})$. We now specialize to a scalar isoscalar
resonance $\sigma$, whose Lagrangian is given by $\LL_\sigma = -
\frac12 \Bigl[ \sigma (M_\sigma^2 + \pd^2) \sigma - g_\sigma v
\tr{\vV_\mu \vV^\mu} - h_\sigma \tr{\vT \vV_\mu} \tr{\vT \vV^\mu}
\Bigr]$. Integrating out the scalar, leads to the effective Lagrangian 
\begin{equation*}
  \LL_\sigma^\eff = \frac{v^2}{8 M_\sigma^2} \biggl[
    g_\sigma \tr{\vV_\mu \vV^\mu} + h_\sigma \tr{\vT
      \vV_\mu} \tr{\vT \vV^\mu} \biggr]^2 \, .
\end{equation*}
From this one can read off that integrating out a scalar isoscalar
generates the following anomalous quartic couplings
\def\fact{\left(\frac{v^2}{8M_\sigma^2}\right)}
\begin{equation}
  \alpha_5 = g_\sigma^2 \fact \, , \qquad
  \alpha_7 = 2 g_\sigma h_\sigma \fact \, , \qquad
  \alpha_{10} = 2 h_\sigma^2 \fact \, .
\end{equation}
One sees immediately, that a heavy SM Higgs would have fit into that
scheme, using the special couplings $g_\sigma = 1$ and $h_\sigma = 0$.

When one tries to turn constraints on anomalous couplings into direct
constraints on new physics, one faces the problem that there are too
many free parameters to overconstrain the system. There is however one
\begin{figure}
  \centering
  \includegraphics[width=.4\textwidth]{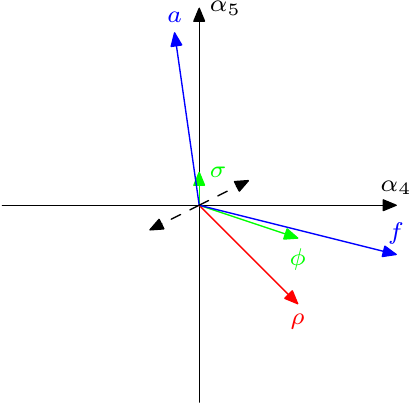}  
  \caption{\label{fig:a4a5} Shifts in the $(\alpha_4,\alpha_5)$ plane 
    through heavy resonances in different spin-isospin
    channels. $\sigma$ and $\phi$ are scalar resonances with $I =
    0,2$, respectively, $\rho$ is a vector isovector, while $f$ and
    $t$ are tensor resonances of $I = 0,2$, respectively. The dashed
    line shows the corrections to the $\alpha$ parameters from higher
    orders in the SM perturbative series.} 
\end{figure}
limiting case where one can do that. This has been applied in the
context of studies of the possible search power of a 1 TeV ILC on
anomalous quartic couplings and their interpretation in terms of
resonances~\cite{Beyer:2006hx}: In the limit of a very broad resonance
(that couples rather strongly to the EW sector), the resonance is
close to a broad continuum: $\Gamma \sim M \gg
\Gamma(\text{non}-WW,ZZ) \sim 0$. Also, in that case the decays of
such a particular resonance into non-$W/Z$s can be ignored. From the
functional relation between the resonance width, its couplings and its
mass (again in the case of a scalar isoscalar)
\begin{equation}
  \Gamma_\sigma = \frac{g_\sigma^2 + \frac12 (g_\sigma^2 + 2
    h_\sigma^2)^2}{16 \pi} \left( \frac{M_\sigma^3}{v^2} \right) + 
  \Gamma (\text{non}-WW,ZZ)
\end{equation} 
one can translate bounds for anomalous couplings directly into
those of the effective Lagrangian:
\begin{equation}
  \alpha_5 \leq \frac{4 \pi}{3} \left( \frac{v^4}{M_\sigma^4}
  \right) \approx \frac{0.015}{(M_\sigma \;\text{in TeV})^4}
  \quad \Rightarrow \quad 16 \pi^2 \alpha_5 \leq
  \frac{2.42}{(M_\sigma \;\text{in TeV})^4} 
\end{equation} 
Note that because of the different dependence of scalar and tensor
widths compared to vector widths, the limits behave differently
depending on the spin of the resonance:
\begin{center}
  \fbox{
    \begin{tabular}{rll}
      Scalar:& $\Gamma \sim g^2 M^3$, $\alpha
      \sim g^2/M^2$ & $\Rightarrow \quad 
        \alpha_{\text{max}}  
        \sim 1 / M^4$   \\
      Vector:& $\Gamma \sim g^2 M$, $\alpha
      \sim g^2/M^2$ & $\Rightarrow \quad 
        \alpha_{\text{max}}  
        \sim 1 / M^2$   \\
      Tensor:& $\Gamma \sim g^2 M^3$,
      $\alpha  
      \sim g^2/M^2$ & $\Rightarrow 
        \alpha_{\text{max}}  
        \sim 1 / M^4$   
    \end{tabular}}
\end{center}
The following table
\begin{equation*}
  \begin{array}{l|ccccc}
    \hline
    \text{Resonance}
    &
    \sigma
    &
    \phi
    &
    \rho
    &
    f
    &
    t
    \\
    \hline
    \Gamma  [g^2 M^2/(64\pi v^2)] 
    &
    6
    &
    1
    &
    \frac{4}{3}(\frac{v^2}{M^2})
    &
    \frac{1}{5}
    &
    \frac{1}{30}
    \\
    \hline
    \\[-9pt]
    \Delta\alpha_4  [(16\pi\Gamma/M)(v^4/M^4)]  &
    0 &
    \frac14 &
    \frac34 &
    \frac52 &
    -\frac58
    \\[6pt]
  \Delta\alpha_5  [(16\pi\Gamma/M)(v^4/M^4)]  &
  \frac{1}{12} &
  -\frac{1}{12} &
  -\frac{3}{4} &
  -\frac{5}{8} &
  \frac{35}{8}
  \\[3pt]
  \hline
  \end{array}
\end{equation*}
shows the width of the five different possible
non-$SU(2)_c$ violating resonances with their widths into longitudinal
EW gauge bosons, as well as their contributions to the anomalous
quartic couplings parameters $\alpha_4$ and
$\alpha_5$. Fig.~\ref{fig:a4a5} shows how the different resonances
would show up in the $(\alpha_4, \alpha_5)$ plane. From this a
discrimination of different resonances even slighly below direct
production of those resonances would be possible.
The dashed lines
in Fig.~\ref{fig:a4a5} show shifts due to higher-order corrections
from SM longitudinal gauge bosons. Note that the treatment of tensor
resonances is notoriously complicated, as their couplings to
longitudinal and transversal currents have to be constructed in
different ways, as will be shown in~\cite{KRS}. 
    
%%%%%%%%%%%%%%%%%%%%%%%%%%%%%%%%%%%%%%%%%%%%%%%%%%%%%%%%%%%%%%%%%%%%%%%%
\section{Vector Boson Scattering at LHC and Unitarity}

In this section, we want to discuss the signatures for vector boson
scattering (VBS) at the LHC as well as issues of perturbative tree-level
\begin{figure}
  \centering
  \includegraphics[width=5cm]{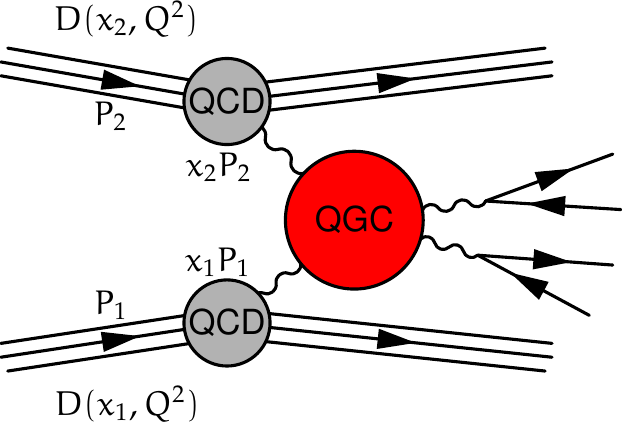}
  \caption{\label{fig:vbs} Signature of vector boson scattering at the
    LHC as a means to measure quartic gauge couplings. } 
\end{figure}
unitarity for our simplified models. At a hadron collider like the
LHC, the typical signature for VBS for measuring (anomalous) quartic
gauge couplings at the LHC is shown in Fig.~\ref{fig:vbs}. When one
takes all leptons (including $\tau$s) as final-state particles, the
cross section at the LHC for $pp \to jj (ZZ/WW) \to jj \ell^- \ell^+
\nu_\ell \bar\nu_\ell$  is roughly $\sigma \approx 40 \,
\text{fb}$. The most severe background comes from top pair production,
$t\bar t\to WbWb$, being three orders of magnitude larger,
$\sigma_{tt} \approx 52 \, \pb$. Also, single top where
one of the jets has been misreconstructed contributes with a cross
section of $\sigma_t \approx 4.8 \, \pb$ to the background, while the
QCD background -- though still sizeable -- is not that bad:
$\sigma_{QCD} \approx 0.21 \,\pb$. To separate VBS from the
background, a tag of two identified and separated leptons and two jets
each is applied, where a large rapidity distance of the two jets is
demanded in order to take into account collinear radiation, 
$|\Delta \eta_{jj}| > 4.4 $, and explicit vetos against $b$ jets. The
leptons should be in a fiducial volume in the central part of the
detector, $\eta_{tag}^{min} < \eta_\ell < \eta_{tag}^{max}$, there is
a lower cut on the dijet invariant mass, e.g. $M_{jj} > 1000 \,\GeV$,
lower cuts on the jet energy (e.g. $E_j > 600 , 400 \,\GeV$) as well
as lower cuts on the two jet $p^1_{T,j} > 60, 24 \, \GeV$ (all values
are just rough estimates). Particularly the large dijet invariant mass
is a powerful means to discriminate against top contamination. At the
moment, it is still unclear whether vetoing against hadronic activity
in the central part of the detector is feasible or not. In general, cuts
like those mentioned help to improve the signal-over-background ration
by roughly one order of magnitude.   

Now, we discuss the issues of perturbative unitarity within one
example of our simplified models, a model that contains a scalar
resonance explicitly, but also anomalous quartic gauge couplings
explicitly. This sounds contrived at first, but such
$\mathcal{O}_{4,5}$ can easily arise through other resonances with
different spin or different recurrences of scalar resonances, e.g. in
extra-dimensional models. Furthermore, they can be generated through
higher-order corrections, which in strongly-interacting models can be
sizeable. Clearly, in nature, unitarity will never be violated, it is
just the truncation of a perturbative series in a simplified model/an
effective field theory that leads to a possible violation of
lowest-order perturbative unitarity. A UV completion of such a model
has to restore unitarity again (possibly through higher orders). 
For the discussion of the issues of unitarity and an algorithm to a
prescription that does not violate unitarity, we review the issue of
perturbative unitarity in Goldstone boson scattering, taking the 
lowest-order EW Lagrangian including the Higgs and anomalous
couplings: 
\begin{equation}
  \LL = - \frac{v^2}{4} \tr{\vV_\mu \vV^\mu}
  + \frac{g_h v}{2} \tr{\mathbf{V}^\mu \mathbf{V}_\mu} h 
  + \alpha_4
  \tr{\vV_\mu\vV_\nu} \tr{\vV^\mu\vV^\nu} + \alpha_5 \left(
    \tr{\vV_\mu \vV^\mu} \right)^2 
\end{equation}
Using the standard Mandelstam variables, $s = (p_1+p_2)^2$, $t = (p_1 
- p_3)^2$ and $u = (p_1-p_4)^2$, this leads to the following
amplitudes for the scattering of longitudinal EW vector bosons:
\begin{subequations}
\label{eq:goldstonescattering}
\begin{align}
  \amp(s,t,u) =: \qquad 
  \amp(w^+w^-\to zz)
  &= 
  \phantom{-}\frac{s}{v^2} - \phantom{\sum_{x=s,t}}
  \frac{g^2_h}{v^2}\frac{s^2}{s-M_H^2} 
  \quad + 8 \alpha_5 \frac{s^2}{v^4} 
  +  4 \alpha_4 \frac{t^2+u^2}{v^4}
  \\
  \amp(w^+z\to w^+ z)
  &=
  \phantom{-} \frac{t}{v^2} - \phantom{\sum_{x=s,t}}
  \frac{g^2_h}{v^2}\frac{t^2}{t-M_H^2} \quad 
  + 8 \alpha_5 \frac{t^2}{v^4}
  + 4 \alpha_4 \frac{s^2 + u^2}{v^4}
  \\
  \amp(w^+w^-\to w^+w^-)
  &=
  -\frac{u}{v^2}  - \sum_{x=s,t} \frac{g^2_h}{v^2}\frac{x^2}{x-M_H^2} \quad
  + \left( 4 \alpha_4 + 2\alpha_5
  \right) \frac{s^2 + t^2}{v^4} 
  + 8 \alpha_5 \frac{u^2}{v^4}
  \\
  \amp(w^+w^+\to w^+w^+)
  &=
  -\frac{s}{v^2}  - \sum_{x=t,u} \frac{g^2_h}{v^2}\frac{x^2}{x-M_H^2}
  \quad 
  + 8 \alpha_4 \frac{s^2}{v^4}
  + 4 \left( \alpha_4 + 2\alpha_5 \right) 
  \frac{t^2+u^2}{v^4}
  \\
  \amp(zz\to zz)
  &= \phantom{-\frac{s}{v^2}} 
    - \sum_{x=s,t,u} \frac{g^2_h}{v^2}\frac{x^2}{x-M_H^2}
    \quad\; +\;
  8 \left(\alpha_4 + \alpha_5 \right)
  \frac{s^2 + t^2 + u^2}{v^4} 
\end{align}
\end{subequations}
The first term in these equations is the so-called low-energy theorem
(LET) that constitutes the scattering of longitudinal gauge bosons
through themselves, the second term comes from exchange of the Higgs
particle (whose coupling in the SM would be $g_h = 1$), while the
other terms originate from the higher-dimensional operators. 

In order to derive the unitarity limits, one has to decompose this
into the corresponding isospin eigenamplitudes according to the
following Clebsch-Gordan decomposition:
\begin{align}
  \amp(I=0) &=\; 3 \amp(s,t,u) + \amp(t,s,u) + \amp (u,s,t)
  \\
  \amp(I=1) &=\; \amp(t,s,u) - \amp (u,s,t)
  \\
  \amp(I=2) &=\; \amp(t,s,u) + \amp (u,s,t)
\end{align}

Following the discussion in~\cite{Lee:1977yc}, the total cross section 
$\sigma = |\mathcal{A}|^2/(64\pi^2 s)$ due to the unitarity of the $S$
matrix has to obey the optical theorem, $\sigma_{\text{tot}} =
\text{Im} \left[\mathcal{A}_{ii}(t=0)\right]/s$, where the Mandelstam
variable is $t = -s (1-\cos\theta)/2$. In order to check the
\begin{figure}
  \centering
  \includegraphics[width=.45\textwidth]{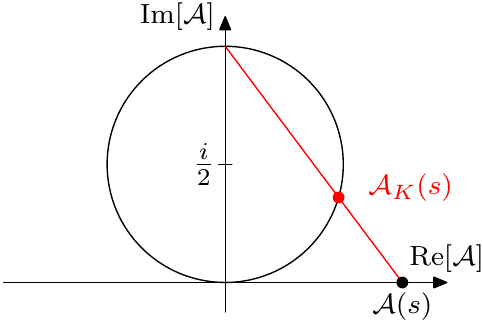}
  \includegraphics[width=.45\textwidth]{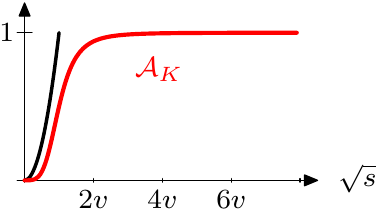}        
  \caption{Left: Argand circle for elastic scattering amplitudes
    according to the optical theorem. Stereographic $K$-matrix
    projection back to the circle for amplitudes violating
    perturbative unitarity. Right: Saturation of an amplitude
    rising quadratically with energy due to $K$-matrix
    unitarization, reaching a constant value.} 
  \label{fig:kmatrix}
\end{figure}
scattering wave unitarity, one has to decompose the quantum-mechanical
amplitude into partial wave amplitudes, $\mathcal{A}(s,t,u) = 32\pi
\sum_\ell (2\ell +1) \amp_\ell(s) P_\ell(\cos\theta)$. Assuming only
elastic scattering processes and demanding the equality between total
cross section and the imaginary part of the forward scattering
amplitude for the partial wave results in the condition: $\left|
\amp(s) - \tfrac{i}{2} \right| = \tfrac{1}{2}$. This means that the
elastic scattering eigenamplitude has to lie on the Argand circle
(Fig.~\ref{fig:kmatrix}, left).  

From the Goldstone scattering amplitudes
Eq.~\ref{eq:goldstonescattering}, project out the isospin
eigenamplitudes~\cite{Lee:1977yc}, which leads (only for the LET-part)
to the three following isopin-spin eigenamplitudes: 
\begin{equation}
  \amp_{I=0 , \,J=0} = \frac{s}{16 \pi v^2} \qquad  
  \amp_{I=1 , \,J=1} = \frac{s}{96 \pi v^2} \qquad  
  \amp_{I=2 , \,J=0} =  - \frac{s}{32 \pi v^2} \; .
\end{equation} 
The condition $|\amp_{IJ}| \lesssim \frac12$ leads to the famous 
unitarity bounds, $E \sim\; \sqrt{\left\{ 8, 48, 16\right\} \pi} v  = 
\left\{ 1.2, 3.5, 1.7 \right\} \,\TeV$, respectively. 
The Higgs exchange (the second term in
Eq.~\ref{eq:goldstonescattering}) ameliorates the quadratic rise in
energy to $\amp(s,t,u) = - (M_H^2/v^2) \times s/(s-M_H^2)$, which
leads to the (tree-level) unitarity bound for a heavy SM Higgs boson
of $M_H \lesssim \sqrt{8 \pi} v \sim 1.2\,\TeV$. 

In the SM with a 125 GeV Higgs boson no problem with perturbative
unitarity should arise, and no deviations should be visible in VBS
from their SM predictions. However, even slight deviations in the
Higgs couplings, $g_h = 1$, would lead to uncanceled unitarity
violation. Also simplified models to test the spin of the 125 GeV
Higgs against are difficult to define in a sane way as one has to take
the scalar Higgs out and introduce a tensor resonance in order to
exclude spin 2 from data. In all such cases, simplified models could
arise that suffer from the issue of perturbative unitarity
violations. Particularly, all simplified models endowed with
resonances motivated by nearly all BSM models mentioned above do
so. In order to get a theoretical description (e.g. for a Monte Carlo
simulation), a prescription that leads to a simplified model covering
the gross features of such BSM models, but on the other hand yielding
amplitudes consistent with unitarity constraints would be highly
welcome. Such an algorithm was proposed in~\cite{Alboteanu:2008my} and
is further refined in~\cite{KRS}. 

One straightforward prescription is the so-called $K$-matrix
unitarization that has been used for the descrition of pion scattering
processes. It consists of using a stereographic projection of an
amplitude exceeding the unitarity constraint on the real axis back to
the Argand circle (cf. left hand side of Fig.~\ref{fig:kmatrix}): 
\begin{equation}
  \amp_K(s) = \frac{\amp(s)}{1 - i \amp(s)} = \amp(s)
  \frac{1 + i \amp(s)}{1 + \amp(s)^2} 
\end{equation} 
Physically, this would correspond to unitarization by an infinitely
heavy and infinitely wide resonance (for more technical details,
references, and also relations to other unitarization prescriptions
cf.~\cite{Alboteanu:2008my,KRS}). This prescription ameliorates
a e.g. quadratic (or quartic) rise of an amplitude to a constant just
saturating the unitarity bound at the very point where unitarity
violation would start to set in (right hand side of
Fig.~\ref{fig:kmatrix}).   

We now show how this unitarization prescription works in the case of a
scalar isosinglet resonance. Here, the Lagrangian including the
kinetic term and the coupling to the current of the longitudinal EW
gauge bosons is given by
\begin{equation}
  \label{eq:scalarsinglet}
  \LL_\sigma = -\tfrac12 \sigma\left(M_\sigma^2 + \pd^2\right)\sigma
  + \tfrac{g_\sigma v}{2} \sigma \tr{\vV_\mu\vV^\mu}   \, , 
\end{equation}
which leads to the following Feynman rules: $\sigma w^+ w^-: \; -
\tfrac{2 i g_\sigma}{v} (k_+ \cdot k_-)$, $\sigma z z: \; - \tfrac{2 i
g_\sigma}{v} (k_1 \cdot k_2)$. Note that this is in complete analogy
to the case of the SM Higgs which has the same quantum numbers, except
that now the mass and the coupling are completely arbitrary. Hence,
the amplitude for the $s$-channel exchange is $\amp^\sigma(s,t,u) = 
(g_\sigma^2 / v^2) \times s^2 / (s-M^2)$. This leads then to the
isospin eigenamplitudes that contain explicit resonance poles, and
spin-isospin eigenamplitudes that are no longer polynomial in the
Mandelstam variables. For e.g.:
\begin{equation}
  \label{eq:a00}
  \amp^\sigma_{00}(s) =\;  3\tfrac{g_\sigma^2}{v^2}
  \tfrac{s^2}{s-M^2} 
  + 2\tfrac{g^2}{v^2} \mathcal{S}_0(s)  \qquad\qquad
  \mathcal{S}_0(s) = M^2 - \tfrac s2 + \frac{M^4}{s} \log
  \frac{s}{s+M^2}  
\end{equation}
For the $K$-matrix unitarization, the $s$-channel resonance pole must
be treated separately in special way. We define the complete
spin-isospin eigenamplitude as the SM amplitude (including the Higgs
boson exchange), $A^{(0)}_{IJ}(s)$, then a BSM contribution (due to
anomalous couplings or finite pieces of resonance exchange, or due to
deviations of the Higgs amplitude from its SM value), $F_{IJ}(s)$, and
the explicit resonance pole, $G_{IJ}(s)$:
\begin{equation}
  A_{IJ}(s) = A^{(0)}_{IJ}(s) + F_{IJ}(s) + \frac{G_{IJ}(s)}{s-M^2}
%   \quad. 
\end{equation}
The coefficient function $G_{IJ}(s) \propto s$ for vector resonances,
and $\propto s^2$ for scalar and tensor resonances, respectively. 
Applying the $K$-matrix projection leads to a correction of the
amplitude, which can be redefined as an additive correction to the 
original amplitude:
\begin{equation}
  \label{eq:kmatrixcorrection}
  \hat A_{IJ}(s) = \frac{A_{IJ}(s)}{1 - \frac{i}{32\pi}A_{IJ}(s)}
  = A^{(0)}_{IJ}(s) + 32 \pi i \Delta A_{IJ}(s),
\end{equation}
with         
\begin{equation}
  \label{eq:correctionterm}
  \Delta A_{IJ}(s) = 32\pi
  i\left(1 \hspace{-0.4mm}+\hspace{-0.4mm}
    \tfrac{i}{32\pi}A^{(0)}_{IJ}(s) 
    \hspace{-0.4mm}+ \hspace{-0.4mm}\frac{s - M^2}
    {\frac{i}{32\pi} G_{IJ}(s)
      - (s - M^2)\left[1 - \frac{i}{32\pi}
        (A^{(0)}_{IJ}(s) + F_{IJ}(s))\right]
    }\right)  
\end{equation}
In order to implement this into a Monte Carlo event generator, one has
to explicitly take care that the unitarization prescription by means
of the $K$-matrix projection only happens in $s$-channel like
configurations. Hence, such an algorithm breaks crossing symmetry. The
formalism described here in form of the simplified models discussed
above has been implemented and validated in the event generator
WHIZARD~\cite{Kilian:2007gr,Moretti:2001zz}. Its setup of the
color-flow formalism~\cite{Kilian:2012pz} as well as the connection to
the parton shower~\cite{Kilian:2011ka} are both compatible with the
formalism of the $K$-matrix prescription. For more technical details
(also a validation using the FeynRules
\begin{figure}
  \centering
  \includegraphics[width=.76\textwidth]{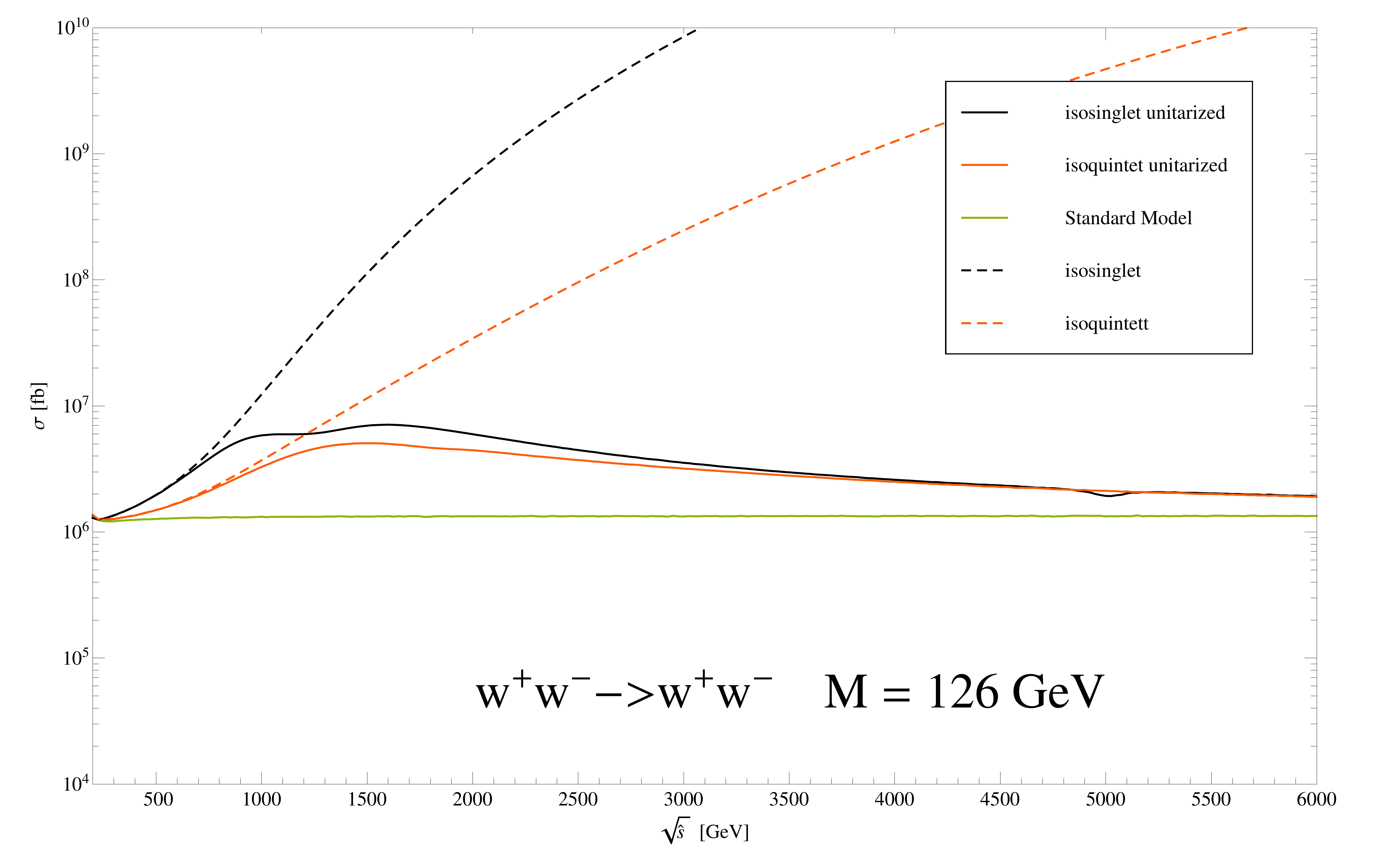}    
  \caption{Scalar resonances the scattering of longitudinal $W$
    bosons, $w^+w^- \to w^+w^-$. The green line is the SM cross
    section which is constant in energy. The dashed black and red
    line show the cross sections for the scalar isosinglet $\sigma$
    and the isotensor $\phi$, respectively. The unitarization
    violation is clearly visible. The full black and red lines are
    the ones with $K$-matrix unitarization.}
  \label{fig:scalarres}
\end{figure}
interface~\cite{Christensen:2010wz}) cf.~\cite{KRS}. 

Using that implementation, the sensitivity of a 1 TeV ILC with 1
ab${}^{-1}$ has been determined in an extensive
study~\cite{Beyer:2006hx}. There, the 1 $\sigma$ sensitivity on the
anomalous couplings $\alpha_4$ and $\alpha_5$ turn out to be 0.0088
and 0.0071, respectively. This translate into the following reach
limits for pure EW resonances in the setup of a 1 TeV ILC (in TeV):
\begin{equation*}
  \begin{array}{|c||c|c|c|}
    \hline
    \text{Spin} & I=0 & I=1 & I =2
    \\
    \hline\hline
    0 & 1.39 & 1.55 & 1.95
    \\
    1 & 1.74 & 2.67 & -
    \\
    2 & 3.00 & 3.01 & 5.84
    \\
    \hline
  \end{array}
\end{equation*}
For this 1 TeV ILC study, possible unitarity violation issues have not
yet played a role, but both 8 and 14 TeV LHC runs as well as a EW
physics at a 3 TeV CLIC have to take this into account. No
concise LHC study has as yet been done for these simplified models,
but more results will be given in~\cite{KRS}. Generically, one can
deduce that the sensitivity on new resonances rises with the number of
intermediate (spin) states, such that tensor resonances have higher
reach than vectors or even scalar resonances. A first rough estimate
from~\cite{Alboteanu:2008my} might serve as a guideline for the
expectations from 300 fb${}^{-1}$ at 14 TeV LHC, where the expected
reach is varying from 0.5 TeV to 2 TeV for scalar up to tensor
resonances. 

%%%%% %%%%% %%%%%

\section{Summary and Conclusion}
\label{sec:conclusion}

In this contribution to the Snowmass White Paper we were constructing
simplified models that allow to describe the physics of a modified
electroweak sector compared to the SM. Experimental observables for
such scenarios are modifications of diboson production, triboson
production, and particularly vector boson scattering. Modifications of
the EW sector could be just parameterized by deviations in the triple
and quartic gauge couplings. In such a case a convenient operator
basis is the one from the EW chiral Lagrangian that has now been
enlarged by all operators containing the Higgs state observed from the
LHC. However, most models have their most natural regions of parameter
space where new resonances show up directly in the upcoming 14 TeV of
LHC. In such a case, a description with higher-dimensional operators
alone is insufficient and not applicable. The simplified models
discussed here contain the SM supplemented by all possible resonances
that could couple to the sector of EW gauge bosons according to their
spin and isospin quantum numbers. Such simplified models cover cases
ranging from Two- or Multi-Higgs doublet models, extended scalar
sectors, Technicolor models, models of complete or partial
compositeness, Little Higgs models, Twin Higgs models and many more. 
Cases where there is only a single resonance present could be
described along these lines as well as cases where there are more
resonances (but maybe only one of them accessible to LHC). The
resonances are just parameterized by their mass, possibly their width,
as well as their couplings to the electroweak sector. As simplified
models are like any effective field theory not UV-complete,
perturbative unitarity of tree-level amplitudes in that setup are not
guaranteed. To give a prescription that can be used by the experiments
in a model-independent setup and does not yield overly optimistic
results due to unphysical amplitude contributions within exclusion
limits, a unitarization formalism has been introduced that projects
back on amplitudes that are genuinely unitary. This is insured by
giving additive corrections to the SM vector boson scattering
augmented by BSM resonances. A simple implementation has been
performed in the event generator WHIZARD~\cite{Kilian:2007gr}. This
proceedings contribution is intended as a first gathering of the
findings in~\cite{KRS} where also all the technical details can be
found. Special emphasis there is also given to tensor resonances that
have not been discussed that extensively in the literature. There,
particularly a careful treatment of subleading terms in longitudinal
and transversal modes of electroweak gauge bosons is crucial.

For a 1 TeV ILC there was an elaborate joint experimental and
theoretical study that determined the ILC search reach for anomalous
quartic couplings and its re-interpretation in terms of
model-independent resonances. The sensitivity rises with number of
intermediate states, from scalars over vectors to tensors. At the
LHC sensitivity limited in pure EW sector, the projected reach might
lie in the range from $0.6 - 2 \,\TeV$ compared to $1 - 6 \,\TeV$.
More studies are urgently needed to find out whether a high-luminosity
phase of the 14 TeV LHC or a higher-energy upgrade are the better
options for these kinds of extensions of the EW sector. Also, it is
not yet clear whether cut-based approaches or multi-variate analyses
give the best sensitivities. More kinematic variables have to be
investigated in order to optimize the reach of the LHC even with only
300 fb${}^{-1}$ for vector boson scattering.

%%%--- Bibliography ---%%%
\bibliographystyle{unsrt}

\end{document}